\documentclass[a4paper,12pt]{article}

\RequirePackage{natbib}
\usepackage{amsmath}
\usepackage{amssymb}
\usepackage{latexsym}
\usepackage{ntheorem}
\usepackage{ifthen}
\usepackage{xcolor}
\usepackage{newcent}
\usepackage[all,cmtip]{xy}

\usepackage{smartdiagram}
\usepackage{centernot}
\usepackage{tikz-cd}
\usepackage {mathtools}
\mathtoolsset{showonlyrefs=true}

\usepackage{setspace}

\usepackage[margin=2.0cm]{geometry}

\theoremstyle{plain}

\theoremheaderfont{\bf}\theorembodyfont{\sl}

\newtheorem{theorem}{Theorem}[section]
\newtheorem{proposition}[theorem]{Proposition}
\newtheorem{lemma}[theorem]{Lemma}

\theoremheaderfont{\it}\theorembodyfont{\rm}

\newtheorem{remark}[theorem]{Remark}

\theoremheaderfont{\bf}\theorembodyfont{\rm}

\newtheorem{definition}[theorem]{Definition}
\newtheorem{example}[theorem]{Example}

\newtheorem{assumption}[theorem]{Assumption}

\theoremstyle{nonumberplain}

\theoremheaderfont{\bf}\theorembodyfont{\sl}

\newenvironment{proof}[1][]
{\ifthenelse{\equal{#1}{}}{\smallskip\noindent\textsl{Proof. }}{\smallskip
\noindent\textsl{Proof #1. }}}{\hfill$\Box$}

\def\Qc{\mathcal{Q}}
\def\Pc{\mathcal{P}}
\def\Bc{{\cal B}}

\def\Pc{{\cal P}}

\def \l{\lambda}

\def \o{\omega}
\def \O{\Omega}

\def \e{\varepsilon}
\begin{document}

\title{Super-replication prices with multiple-priors in discrete time}

\author {
{Romain} {Blanchard}, E.mail~:  romblanch@hotmail.com\\
\and
{Laurence} {Carassus}, E.mail~: laurence.carassus@devinci.fr \\
L\' eonard de Vinci P\^ole Universitaire, Research Center, \\
92916 Paris La D\'efense, France and 
 \\
LMR, FRE 2011 Universit\'e Reims Champagne-Ardenne.\\
}

\maketitle

% REQUIRED
\begin{abstract}
 In the frictionless discrete time financial market of Bouchard and Nutz (2015), we propose a full characterization of the quasi-sure super-replication price: as the supremum of  the mono-prior super-replication prices, through an extreme prior and through martingale measures. 
\end{abstract}

% REQUIRED

% REQUIRED

\section{Introduction}
We work in the  discrete time and multiple-priors setting of \cite{BN} and consider an abstract  set of priors $\Qc$, possibly large and in particular not dominated by a single probability measure or weakly compact, see \citep{Bart16, BC19} for concrete examples including  non-dominated settings. 
We consider  an agent who want to price a contingent claim $H$ maturing at a future date $T$ in a robust and risk-conservative way and compute for that  the multiple-prior super-hedging price $\pi^{\mathcal{Q}}(H),$ i.e. the price of cheapest trading strategies which are guaranteed to cover  $H$ under all of priors $\Qc.$ 
The super-hedging price $\pi^{\mathcal{Q}}(H)$ can be characterised theoretically and has been considered in a number of papers, see among others  \cite{BN} and  \cite{COW}.
Here, we introduce  $\underline{\pi}^{\mathcal{Q}}(H),$ which corresponds to the pricing/hedging problem where the agent want to find a super-hedging strategy for each prior in $ \mathcal{Q}$ and then receive the infimum wealth such that this is possible. This is in contrast with $\pi^{\mathcal{Q}}(H)$ which corresponds to the case where the same super-hedging strategy must hold for all the priors.  Clearly, $\underline{\pi}^{\mathcal{Q}}(H)$ should provide a cheaper alternative to the classical multiple-prior super-hedging price. \\
Theorem \ref{super} is our main result. We prove that for some set of priors $\Pc$ satisfying Assumption \ref{PaQ}, 
%(\red{which turns to be equivalent to the quasi-sure no-arbitrage condition}) 
$\pi^{\mathcal{Q}}(H)$, $\pi^{\mathcal{P}}(H)$ and $\underline{\pi}^{\mathcal{P}}(H)$ are equal. We also provide some dual characterizations through martingale measures and show that $\pi^{\mathcal{Q}}(H)$ is the supremum  over all the priors of $\mathcal{P}$ of the mono-prior super-replication  prices. 
We prove that this is also the case for $\mathcal{Q}$ in Proposition \ref{pasuper}, where we show as well that $\pi^{\mathcal{Q}}(H)$ and  $\underline{\pi}^{\mathcal{Q}}(H)$ are actually equal.  
Finally, we  establish that there exists some prior $\widehat P$ satisfying no-arbitrage and such that the mono-prior price $\pi^{\widehat P}(H)$  equals the multi-prior prices. Our result related to $\widehat P$ is similar to the one obtained in \cite{OW} but our proof relies only on primal arguments. We provide a counter-example to $\widehat P \in \Pc$ as well as condition for this to hold true. \\
We apply Theorem \ref{super}  to the set of priors $\overline{\Pc} \subset \Qc$  obtained in \citep[Theorem 3.6]{BC19} under the quasi-sure no-arbitrage condition. Under the same assumption, we also apply it to a set of priors  $\widetilde{\Pc},$ which is not a priori included in $\Qc,$ but have the same support as  $\Qc.$ This allows  to get a measure $\widehat P$ having also the same support. To do that, we construct by measurable selection arguments a measure $\widetilde P $ having the same support as $\Qc,$ see Lemma \ref{lemmptilde}. This construct has its  own interest. 

We finish this introduction with some notations and reminders. 
For any Polish space $E,$ 
%(i.e.,  complete and separable metric space) 
we denote by $\mathcal{B} (E)$ its Borel sigma-algebra  and by $\mathfrak{P}(E)$ the set of all probability measures on $(E,\mathcal{B} (E))$. 
%We recall  that $\mathfrak{P}(E)$ endowed with the weak topology is a Polish space (see  \cite[Propositions 7.20 p127, 7.23 p131]{bs}). 
The universal sigma-algebra is defined by $\mathcal{B}_{c}(E):= \bigcap_{P \in \mathfrak{P}(E)} \mathcal{B}_{P}(E),$ where $\mathcal{B}_{P}(E)$ is the completion of $\mathcal{B}(E)$ with respect to $P$ in $\mathfrak{P}(E)$. 
%In the rest of the paper, we will use the same notation for $P$ in $\mathfrak{P}(E)$ and for its (unique) extension on $\mathcal{B}_{P}(E)$. 

An analytic set of $E$ is the continuous image of a Polish space, see  \cite[Theorem 12.24 p447]{Hitch}. Moreover, a function $f: E  \to \mathbb{R} \cup \{\pm \infty\}$ is  upper-semianalytic (usa) if  for all $c \in \mathbb{R},$ $\{x \in E,\; f(x)>c\}$ is an analytic set,  see   \cite[Definition 7.21 p177]{bs}. 

For a given $\mathcal{P} \subset  \mathfrak{P}(E)$, 
a set $N \subset E$ is called a $\mathcal{P}$-polar set if there exists some $A \in \mathcal{B}(E)$ such that $N \subset A$ and $P(A)=0$ for all $P \in \mathcal{P}$. 
%\red{Def de BN, Nutz, Burzoni Fritteli Hou Maggis et Obloj, means negligible for the capacity $c(A)=\sup_{P \in \Qc^t} P(A)$ It is not our former def and not the one of Burzoni Maggis, Berhektiar Zou}. 
A property holds true $\mathcal{P}$-quasi-surely (q.s.), if it is true outside a $\mathcal{P}$-polar set and a set is of $\mathcal{P}$-full measure  if its complement is a $\mathcal{P}$-polar set.

The remainder of the paper is organised as follows. The next section introduces and discusses our modelling framework.  Section \ref{main} presents our main theorem 
while Section \ref{appli} presents the applications

\section{Uncertainty modelling}
We fix  a time horizon $T\in \mathbb{N}$ and  introduce  a sequence $\left(\Omega_t\right)_{1 \leq t \leq T}$  of Polish spaces.    
 For some $1 \leq t \leq T$, we denote by $\Omega^{t}:=\O_{1} \times \dots \times \O_{t}$  (with the convention that $\Omega^{0}$ is reduced to a singleton).  
 %An element of $\Omega^{t}$ will be denoted by $\o^{t}=(\omega_{1},\dots, \omega_{t})=(\o^{t-1},\o_{t})$ for $(\o_{1},\dots,\o_{t}) \in \Omega_{1}\times\dots\times\Omega_{t}$. \\
Let $S:=\left\{S_{t},\ 0\leq t\leq T\right\}$ be a 
$\mathbb{R}^d$-valued process, 
where for all $0 \leq t \leq T$, $S_{t}=\left(S^i_t\right)_{1 \leq i \leq d}$ represents the   price of $d$ (discounted) risky assets at time $t$.   The notation $\Delta
S_t:=S_t-S_{t-1}$ will often be used. 
\begin{assumption}
\label{SassARB}
The process $S$ is $\left(\mathcal{B}(\Omega^{t})\right)_{0 \leq t \leq T}$-adapted.
 \end{assumption}
Let $\phi:=\{ \phi_{t}, \,{1 \leq t \leq T}\},$ where for all $1 \leq t \leq T$, $\phi_{t}=\left(\phi^{i}_{t}\right)_{1 \leq i \leq d}$ represents the
investor's holdings in  each of the $d$ assets at time $t$.  
We assume that $\phi$ is $\left(\mathcal{B}_{c}(\Omega^{t-1})\right)_{1 \leq t \leq T}$-measurable. Trading is also assumed to be self-financed and the set of such trading strategies is denoted by $\Phi$. The value at time $t$ of a portfolio $\phi \in \Phi$ starting from
initial capital $x\in\mathbb{R}$ is given by
$
V^{x,\phi}_t=x+\sum_{s=1}^t  \phi_s \Delta S_s,
$
where if $x,y\in\mathbb{R}^d$ then
the concatenation $xy$ stands for their scalar product. The symbol $|\cdot|$ denotes the Euclidean norm
on $\mathbb{R}^d$ (or on $\mathbb{R})$.

The uncertainty is measured by $\left(\mathcal{Q}_{t}\right)_{1\leq t\leq T}$, where $\mathcal{Q}_{1}$ is a non-empty and convex subset of $\mathfrak{P}(\O_{1})$ and 
 for all $1\leq t\leq T-1$, $\mathcal{Q}_{t+1}: \Omega^t \twoheadrightarrow \mathfrak{P}(\O_{t+1})$  is a random set \footnote{The notation $\twoheadrightarrow$ stands for set-valued mapping.}  satisfying Assumption \ref{QanalyticARB} below. The set $\mathcal{Q}_{t+1}(\o^{t})$  can be seen as the set of all possible  priors for the $t$-th period given the state $\o^{t}$ until time $t$. 
\begin{assumption}
\label{QanalyticARB}
For all $1\leq t\leq T-1$,  $\mathcal{Q}_{t+1}$ is a non-empty and convex valued random set such that 
$
\mbox{graph}(\mathcal{Q}_{t+1}):=\left\{(\omega^{t},p) \in \Omega^{t}\times \mathfrak{P}(\Omega_{t+1}),\; p \in \mathcal{Q}_{t+1}(\omega^{t})\right\} $ 
is an analytic set.
\end{assumption}
For all $1\leq t\leq T-1$, let $\mathcal{SK}_{t+1}$ be the set of universally-measurable stochastic kernel on $\O_{t+1}$ given $\O^{t}$ (see \citep[Definition 7.12  p134, Lemma 7.28 p174]{bs}). Then, the Jankov-von Neumann Theorem (see  \cite[Proposition 7.49 p182]{bs}) implies that there exists $q_{t+1} \in \mathcal{SK}_{t+1}$ such that for all $\o^{t} \in \O^{t}$, $q_{t+1}(\cdot,\o^{t}) \in \mathcal{Q}_{t+1}(\o^{t})$.  
We set $\mathcal{Q}^{1}=\mathcal{Q}_{1}$ and for all $1\leq t \leq T,$ $\mathcal{Q}^{t} \subset \mathfrak{P}\left(\Omega^{t}\right)$ is defined by
\begin{footnotesize}
\begin{align}
\label{QstarARB}
 \mathcal{Q}^{t}:=\bigl\{ Q_{1} \otimes q_{2} \otimes \dots \otimes q_{t},&\;  Q_{1} \in \mathcal{Q}_{1},\;  q_{s+1} \in \mathcal{SK}_{s+1},  q_{s+1}(\cdot,\o^{s}) \in \mathcal{Q}_{s+1}(\o^{s}),\; \forall \,  \o^{s} \in \O^{s}, \; \forall \,  1\leq s\leq t-1\; \bigr\},
\end{align}
\end{footnotesize}
where $Q^{t}:=Q_{1} \otimes q_{2} \otimes \dots \otimes q_{t}$ denotes the $t$-fold application of Fubini's Theorem (see  \citep[Proposition 7.45 p175]{bs}) which defines a measure on $\mathfrak{P}\left(\Omega^{t}\right).$ We also set $\mathcal{Q}:=\mathcal{Q}^{T}$.

\section{No-arbitrage conditions and super-replication prices}
\label{main}

We recall the following  definitions of no-arbitrage.  The first one is the quasi-sure no-arbitrage of \cite{BN}, while the second one is used in \cite{BC19}. 
\begin{definition}
Let $\mathcal{T} \subset \mathfrak{P}(\O^{T}).$
\begin{itemize}
\item The $NA(\mathcal{T})$ condition holds true if
$V_{T}^{0,\phi} \geq 0 \; \mathcal{T}\mbox{-q.s. for some $\phi  \in \Phi$}$ implies that $ V_{T}^{0,\phi}  = 0 \;\mathcal{T}\mbox{-q.s. }$
\item The  $sNA(\mathcal{T})$ condition holds true if  the $NA(P)$ holds true for all $P \in \mathcal{T}$: i.e., $V_{T}^{0,\phi} \geq 0 \; {P}\mbox{-a.s. for some $\phi  \in \Phi$}$ implies that $ V_{T}^{0,\phi}  = 0 \;P\mbox{-a.s. }$
\end{itemize}
\end{definition}
Let  $H: \O^T \to \mathbb{R}$.  
The multiple-prior super-hedging price of $H$ relative to the set of priors $\mathcal{T} \subset \mathfrak{P}(\O^{T})$  is   given by 
\begin{align}
\label{piM}
\pi^{\mathcal{T}}(H):=\inf \{x \in \mathbb{R}, \; \exists \phi \in \Phi, \; V_{T}^{x,\phi} \geq H\ \mathcal{T} \mbox{-q.s.}\}.
\end{align}
When $\mathcal{T}=\{P\}$ is reduced to a singleton, we note $\pi^{P}(H)$ for the mono-prior superhedging price of $H$ (under $P$). We also define
$$\underline{\pi}^{\mathcal{T}}(H):=\inf \{x \in \mathbb{R}, \forall P \in \mathcal{T}, \; \exists \phi \in \Phi, \; V_{T}^{x,\phi} \geq H\  P \mbox{-a.s.}\}.$$
This last price is smaller than $\pi^{\mathcal{T}}(H)$ as the infimum is taken over a larger set. 
%bIt corresponds to the pricing/hedging problem where the agent want to find a super-hedging strategy for each prior $P \in \mathcal{T}$ and then receive the infimum wealth such that this is possible. The price $\pi^{\mathcal{T}}(H)$ corresponds to the case where the same super-hedging strategy must hold for all the priors of $\mathcal{T}$. 
We also introduce
\begin{align}
\label{Mart0}
M(P) & := \{M \in \mathfrak{P}(\O^{T}),\; M \sim P \; \mbox{and $M$ is a martingale measure}\} \mbox{ for }P \in \mathfrak{P}(\O^{T}) \\
\label{Mart2}
\mathcal{M}_{e,\mathcal{T}}& :=\cup_{Q\in \mathcal{T}} M(Q)\\
\label{Mart}
\mathcal{M}_{a,\mathcal{T}}& :=\{M \in \mathfrak{P}(\O^{T}),\; \exists \, Q \in \mathcal{T}, M \ll Q \; \mbox{and $M$ is a martingale measure}\}.
\end{align}
We now introduce the assumption needed on a set of priors  $\mathcal{P} \subset \mathfrak{P}(\O^{T})$  to obtain Theorem \ref{super}. 
Note that we don't need that $\mathcal{P} \subset \mathcal{Q}$. We will propose in Section \ref{appli} two sets  of priors $\mathcal{P}$ satisfying Assumption \ref{PaQ}.  
\newpage
\begin{assumption}
\label{PaQ}
 {There exists some set   $\mathcal{P} \subset \mathfrak{P}(\O^{T})$ such that }
\begin{itemize}
\item[i)] $sNA(\mathcal{P})$ holds true. 
\item[ii)]  The $\mathcal{Q}$-polar sets are also $\mathcal{P}$-polar sets.
\item[iii)] 
For all $Q \in \mathcal{Q},$ there exists $P\in  \mathcal{P}$ such that $Q \ll  P.$ 
\end{itemize}
 \begin{remark}
 \label{lemem}
 Under Assumption \ref{PaQ} $ii)$ and $iii)$,  $\mathcal{Q}$ and  $\mathcal{P}$ have the same polar sets. As $ii)$ holds true, it is enough to show that any $\mathcal{P}$-polar set $N$ is also a $\mathcal{Q}$-polar set. Let $A \in \Bc(\O^T)$ such that $N \subset A$ and $P(A)=0$ for all $P\in  \mathcal{P}.$ Let $Q \in \mathcal{Q}$. Then, $iii)$ implies that there exists $P\in  \mathcal{P}$ such that $Q \ll  P.$ As $P(A)=0$,  $Q(A)=0$ and $A$ is also a $\mathcal{Q}$-polar set. \\
 Assumption \ref{PaQ} $i)$ implies $NA(\mathcal{P})$ and as $\mathcal{Q}$ and  $\mathcal{P}$ have the same polar sets, $NA(\mathcal{Q})$ also holds true. 
We show in Theorem \ref{spa} that the  $NA(\mathcal{Q})$ condition implies that there exists some set $\overline{\mathcal{P}} \subset \mathcal{Q}$ satisfying Assumption \ref{PaQ}. 
%\red{Theorem \ref{spa} shows that the  $NA(\mathcal{Q})$ condition implies that  Assumption \ref{PaQ} holds true for some set $\overline{\mathcal{P}}\subset \mathcal{Q}$.  Recalling Remark \ref{lemem}, this proves that both conditions are equivalent.}
% Assume that  $NA(\mathcal{Q})$ holds true. 
%  \\
% \red{ REM: If $sNA(\mathcal{P})$ holds true then we already know that $NA(\mathcal{P})$ holds true and we could thus deduce that $NA(\mathcal{Q})$ hold true}
 \end{remark}
\end{assumption}
The following theorem is the main result of the paper.  
%The following theorem proves that under assumption \ref{PaQ},  $\pi^{\mathcal{Q}}(H)$, $\pi^{\mathcal{P}}(H)$ and $\underline{\pi}^{\mathcal{P}}(H)$ are equal. It also give some dual characterizations through $\mathcal{M}_{a,\Qc}$ and $\mathcal{M}_{e,\Pc}.$ It shows that $\pi^{\mathcal{Q}}(H)$ is the supremum  over all the priors of $\mathcal{P}$ of all the mono-prior superreplication  prices. Finally, it also proves that there exists some $\widehat P \in \mathfrak{P}(\O^{T})$ such that the mono prior price $\pi^{\widehat P}(H)$  equals the multi-prior prices. 
\begin{theorem}
\label{super}
Assume that Assumptions  \ref{SassARB} and \ref{QanalyticARB} hold true, as well as Assumption \ref{PaQ}  for some set $\mathcal{P} \subset \mathfrak{P}(\O^{T})$. Let $H: \O^T \to \mathbb{R}$ be usa. Then, 
there exists some $\widehat P \in \mathfrak{P}(\O^{T})$ such that NA$(\widehat P)$ holds true and 
 %$\pi^{\mathcal{Q}}(H)=\pi^{\mathcal{P}}(H) \in (-\infty,\infty)$, $\underline{\pi}^{\mathcal{P}}(H) \in (-\infty,+\infty)$ and
\begin{small}
\begin{align}
\label{price}
-\infty< \pi^{\mathcal{Q}}(H)=\pi^{\mathcal{P}}(H)=\underline{\pi}^{\mathcal{P}}(H)=\sup_{P \in \mathcal{P}} \pi^{P}(H)=\pi^{\widehat P}(H)=\sup_{M \in \mathcal{M}_{e,\Pc}} E_{M} (H)=\sup_{M \in \mathcal{M}_{a,\Qc}} E_{M} (H).  %=E_{P*}(R).
\end{align}
\end{small}
\end{theorem}
\begin{proof} In order to get finite expectations, we proceed as in \citep[Remark 5.2]{BN}. First, note that 
$|H| \leq \varphi :=1+|H|$  and let 
\begin{align}
M_{\varphi}(P) & := \{M \in M(P), \, E_M(\varphi)<\infty \}  \quad 
\mathcal{M}_{\varphi,e,\mathcal{T}}  :=  \{M \in \mathcal{M}_{e,\mathcal{T}}, \, E_M(\varphi)<\infty \} \\
\mathcal{M}_{\varphi,a,\mathcal{T}} & :=  \{M \in \mathcal{M}_{a,\mathcal{T}}, \, E_M(\varphi)<\infty \}. 
\end{align}
Then, all the expectations in \eqref{price} are well-defined and finite considering $\mathcal{M}_{\varphi,e,\Pc}$ instead of $\mathcal{M}_{\varphi,\Pc}$  and $\mathcal{M}_{\varphi,a,\Qc}$ instead of $\mathcal{M}_{a,\Qc}$. Moreover, \citep[Lemma A.3]{BN} shows that 
$\sup_{M \in \mathcal{M}_{e,\Pc}} E_{M} (H)=\sup_{M \in \mathcal{M}_{\varphi, e,\Pc}} E_{M} (H)$
and the same holds true for $\mathcal{M}_{a,\Qc}$. So, instead of \eqref{price}, we prove that 
\begin{small}
\begin{align}
\label{pricephi}
-\infty< \pi^{\mathcal{Q}}(H)=\pi^{\mathcal{P}}(H)=\underline{\pi}^{\mathcal{P}}(H)=\sup_{P \in \mathcal{P}} \pi^{P}(H)=\sup_{M \in \mathcal{M}_{\varphi, e,\Pc}} E_{M} (H)=\sup_{M \in \mathcal{M}_{\varphi, a,\Qc}} E_{M} (H).
\end{align}
\end{small}
%\red{Montrer qui si un vaut $+\infty$ alors tout $+\infty$}.
As both  $NA(\mathcal{Q})$ and  $NA(\mathcal{P})$ hold true (see Remark \ref{lemem}),  using \citep[Theorem 2.3]{BN}, $\pi^{\mathcal{Q}}(H)$ and $\pi^{\mathcal{P}}(H)$ are well-defined and are strictly greater than $-\infty$. The equality $\pi^{\mathcal{Q}}(H)=\pi^{\mathcal{P}}(H)$ follows from the fact that  $\mathcal{P}$ and $\mathcal{Q}$ have the same polar sets. Moreover, using \citep[Theorem 4.9]{BN},  we get that 
\begin{align}
\label{petq}\sup_{M \in \mathcal{M}_{\varphi,a,\Pc}} E_{M}(H)={\pi}^{\mathcal{P}}(H)={\pi}^{\mathcal{Q}}(H)= \sup_{M \in \mathcal{M}_{\varphi,a,\Qc}} E_{M}(H).
\end{align}
Now, we fix some  $P \in \mathcal{P}$.  Since the $sNA(\mathcal{P})$ condition holds true,  using the classical mono-prior  Super-hedging Theorem with the additional weight function $\varphi$ (see  \citep[Proof of Lemma A.3]{BN}),  we get that  $\pi^{P}(H)>-\infty$   and  $\pi^{P}(H)= \sup_{M \in M_{\varphi}(P)} E_{M}(H)$.  Then,  % the set   $\{x \in \mathbb{R}, \forall P \in \mathcal{P}, \; \exists \phi \in \Phi, \; V_{T}^{x,\phi} \geq H \;  P \mbox{-as.}\}$ is not empty and that 
\begin{align}
\label{tord}
\sup_{P \in \mathcal{P}}\pi^{P}(H)&=\sup_{P \in \mathcal{P}} \sup_{M \in M_{\varphi}(P)} E_{M}(H)
= \sup_{M \in \mathcal{M}_{\varphi,e,\Pc}} E_M(H),
\end{align}
and the fourth equality of  \eqref{pricephi} is proved. Now, we prove that 
\begin{align}
\label{tordtord}
\underline{\pi}^{\mathcal{P}}(H)=\sup_{P \in \mathcal{P}} \pi^{P}(H)={\pi}^{\mathcal{P}}(H).
\end{align} 
Fix some $\varepsilon>0$ and  $P \in  \mathcal{P}$. 
There exists some $\phi_P$ such that $V_{T}^{\underline{\pi}^{\mathcal{P}}(H)+ \varepsilon,\phi_P} \geq H\  P \mbox{-a.s.}$ So,  $\pi^{P}(H) \leq \underline{\pi}^{\mathcal{P}}(H)+ \varepsilon$ and $\pi^{P}(H) \leq \underline{\pi}^{\mathcal{P}}(H)$  follows. As this is true for all $P \in \mathcal{P}$, we get that $ \sup_{P \in \mathcal{P}} \pi^{P}(H) \leq \underline{\pi}^{\mathcal{P}}(H)$. 
Now, for all $P \in  \mathcal{P},$ there exists some $\phi_P$ such that $V_{T}^{ \sup_{P \in \mathcal{P}} \pi^{P}(H)+\varepsilon,\phi_{P}} \geq H$ $P$-a.s. and  we get that $\underline{\pi}^{\mathcal{P}}(H) \leq \sup_{P \in \mathcal{P}} \pi^{P}(H)+ \varepsilon.$ Thus, the first equality in \eqref{tordtord} is proved. 
For the second one, the fact that  $\underline{\pi}^{\mathcal{P}}(H) \leq {\pi}^{\mathcal{P}}(H)$ is clear. To show the other inequality, we prove below that $\sup_{M \in \mathcal{M}_{\varphi,a,\Qc}} E_{M}(H) \leq \sup_{M \in \mathcal{M}_{\varphi,e,\Pc}} E_{M}(H)$. This will imply, recalling \eqref{petq}, \eqref{tord} and the first equality in  \eqref{tordtord}, that 
$${\pi}^{\mathcal{P}}(H)=\sup_{M \in \mathcal{M}_{\varphi,a,\Qc}} E_{M}(H) \leq \sup_{M \in \mathcal{M}_{\varphi,e,\Pc}} E_{M}(H)=\sup_{P \in \mathcal{P}}\pi^{P}(H)=\underline{\pi}^{\mathcal{P}}(H).$$ 
We fix some $M\in \mathcal{M}_{\varphi,a,\Qc}$.  Recalling \eqref{Mart}, there exists some $Q \in \mathcal{Q}$ 
 such that $M \ll Q$, $M$ is a martingale measure and $E_M (\varphi )< \infty$.  
Using Assumption \ref{PaQ} $iii)$,  there exists  ${P} \in  \mathcal{P}$ such that $Q \ll  {P}$ and thus $M \ll {P}.$ 
As ${P}\in \mathcal{P}$,  Assumption \ref{PaQ} $i)$ and \citep[Lemma A.3]{BN}) establish the existence of some $\widehat{M} \sim {P}$ such that $\widehat{M}$ is a martingale measure and $E_{\widehat{M}} (\varphi) < \infty,$ i.e., $\widehat{M} \in \mathcal{M}_{\varphi,e,\Pc}$.   For some $n \geq 1,$ set 
$M_{n}:=\left(1-\frac1n\right)M+\frac1n \widehat{M}.$ 
Then, $M_{n}$ is a martingale measure and ${M}_{n} \sim {P},$ so,  ${M}_{n} \in \mathcal{M}_{\varphi,e,\Pc}$.  Indeed, if $M_{n}(A)=0$, then $ \widehat{M}(A)=0$ and ${P}(A)=0$ also since $\widehat{M} \sim {P}$. Now, if ${P}(A)=0$,   $ \widehat{M}(A)=0$ and since $M \ll {P}$, $M(A)=0$  and also $M_{n}(A)=0.$ 
So, 
$$\left(1-\frac1n\right) E_{M} H  + \frac1n E_{\widehat{M}} H=E_{M_{n}} H  \leq \sup_{P \in \mathcal{M}_{\varphi,e,\Pc}} E_{P}(H).$$
Letting $n$ go to $\infty$, we obtain that $E_{M} H \leq \sup_{P \in \mathcal{M}_{\varphi,e,\Pc}} E_{P}(H).$ As this is true for all $M \in \mathcal{M}_{\varphi,a,\Qc}$, we get that $\sup_{M \in \mathcal{M}_{\varphi,a,\Qc}} E_{M}(H) \leq \sup_{M \in \mathcal{M}_{\varphi,e,\Pc}} E_{M}(H)$.\\
%as $ \mathcal{M}_{\varphi,e,\Pc} \subset 
%\mathcal{M}_{\varphi, a,\Pc}$ and recalling \eqref{petq} and \eqref{tord}, we first get that 
%\begin{align}
%\label{ilpleut}
%{\pi}^{\mathcal{P}}(H)=\sup_{M \in \mathcal{M}_{\varphi,a,\Pc}} E_{M}(H) \geq \sup_{M \in \mathcal{M}_{\varphi,e,\Pc}} E_{M}(H)=\sup_{P \in \mathcal{P}}\pi^{P}(H)=\underline{\pi}^{\mathcal{P}}(H).
%\end{align}
Finally, we construct some  $\widehat{P}$ such that the fourth equality in \eqref{price} as well as $NA(\widehat{P})$ holds true. Let $(P_n)_{n \geq 1} \subset \mathcal{P}$  such that $\sup_{P \in \mathcal{P}}\pi^{P}(H)=\sup_{n \geq 1}\pi^{P_n}(H)$. Let \begin{align}
\label{Phat}
\widehat{P}  :=\sum_{n \geq 1} 2^{-n}P_n.
\end{align}
%\begin{align}
%\label{chapeau}
%\widehat{P} & :=\sum_{n \geq 1} 2^{-n}P_n. 
%\end{align}
We prove that \begin{align}
\label{chapeaugale}
\pi^{\mathcal{P}}(H) & =\pi^{\widehat{P}}(H).
\end{align}
As for all $n \geq 1$, $P_n \ll  \widehat{P}$, we get that $\pi^{{P_n}}(H) \leq \pi^{\widehat{P}}(H)$ and using \eqref{tordtord}, 
${\pi}^{\mathcal{P}}(H) \leq \pi^{\widehat{P}}(H).$
Now, we remark that 
$ \{x \in \mathbb{R}, \; \exists \phi \in \Phi, \; V_{T}^{x,\phi} \geq H\ \mathcal{P} \mbox{-q.s.}\} \subset 
 \{x \in \mathbb{R}, \; \exists \phi \in \Phi, \; V_{T}^{x,\phi} \geq H\ \widehat{P}\mbox{-a.s.}\}$ and \eqref{chapeaugale} follows. 
 Indeed, if $V_{T}^{x,\phi} \geq H\ \mathcal{P} \mbox{-q.s.}$ there exists some $A \in \Bc(\O^T)$ such that $A \subset \{V_{T}^{x,\phi} \geq H\}$ and $P(A)=1$ for all $P\in 
\mathcal{P}$. Thus, $P_n(A)=1$ and also $\widehat{P}(A)=1.$ We prove that NA$(\widehat{P})$ holds true. Let $\phi \in \Phi$ such that $V_{T}^{x,\phi} \geq 0\ \widehat{P}\mbox{-a.s.}$ As for all $n \geq 1$, $P_n \ll  \widehat{P}$, $ V_{T}^{x,\phi} \geq 0\ {P}_n\mbox{-a.s.}$ Since $(P_n)_{n \geq 1} \subset \mathcal{P},$ Assumption \ref{PaQ} $i)$ 
shows that $P_n(\{V_{T}^{x,\phi} =0\})=1$ and also $\widehat{P}(\{V_{T}^{x,\phi} = 0\})=1$ and this concludes the proof.\end{proof}

\newpage

\section{Applications}
\label{appli}
We propose a first application of Theorem  \ref{super}.  
\begin{theorem}
\label{spa}
Assume that Assumptions  \ref{SassARB}, \ref{QanalyticARB} and that the $NA(\mathcal{Q})$ condition hold true.  
Then, there exists some set $\overline{\mathcal{P}} \subset \mathcal{Q}$ satisfying Assumption \ref{PaQ}. Assume that $H: \O^T \to \mathbb{R}$ is usa. 
%there exists some $\overline{\mathcal{P}} \subset \mathcal{Q}$  such that  $\overline{\mathcal{P}}$   and $\mathcal{Q}$ have  the  same polar sets and that  the $sNA(\overline{\mathcal{P}})$ condition holds true. 
There exists some $\widehat P \in \mathfrak{P}(\O^{T})$ such that NA$(\widehat P)$ holds true and 
\begin{small}
\begin{align}
\label{price2}
-\infty<\pi^{\mathcal{Q}}(H)=\pi^{\overline{\mathcal{P}}}(H)=\underline{\pi}^{\overline{\mathcal{P}}}(H)=\sup_{P \in \overline{\mathcal{P}}} \pi^{P}(H)=\pi^{\widehat P}(H)=\sup_{R \in \mathcal{M}_{e,\overline{\mathcal{P}}}} E_{R} (H)=\sup_{R \in \mathcal{M}_{a,\Qc}} E_{R} (H). 
\end{align}
\end{small}
\end{theorem}
\begin{proof}
First, $i)$ and $ii)$ in Assumption \ref{PaQ}  hold true for the set $\overline{\mathcal{P}}$ constructed in \citep[Theorem 3.6]{BC19} under $NA(\mathcal{Q})$. Moreover, \citep[Lemma 3.9]{BC19} shows that 
 for all  $Q \in \mathcal{Q}$, there exist some $(R_{k})_{0 \leq k \leq T-1} \subset \mathcal{Q}$ such that  
${P}:=\frac{1}{2^T}Q+ \frac1{2^T}\sum_{k=0}^{T-1} R_{k} \in  \overline{\mathcal{P}}.$
So,  $Q \ll  P$ and $iii)$ in Assumption \ref{PaQ}  is proved. 
Then, we apply  Theorem \ref{super} to $\overline{\mathcal{P}}$. 
\end{proof}
\begin{remark}
The set $\overline{\mathcal{P}}$ in \eqref{price2} can be replaced by any set of priors included in $\mathcal{Q},$ which satisfies Assumption \ref{PaQ}. This is the case for the set $\mathcal{Q}^{*}:=\{Q \in \mathcal{Q},\; \mbox{NA(Q) holds true}\},$ which contains  $\overline{\mathcal{P}}$. %The fact that  $\pi^{\mathcal{Q}}(H)=\underline{\pi}^{{\mathcal{Q}}}(H)$
\end{remark}

\begin{proposition}
\label{pasuper}
Assume that Assumptions  \ref{SassARB} and \ref{QanalyticARB} hold true, as well as Assumption \ref{PaQ}  for some set $\mathcal{P} \subset \mathfrak{P}(\O^{T})$. Let $H: \O^T \to \mathbb{R}$ be usa. Then,  we also have that
\begin{small}
\begin{align}
\label{price3}
-\infty< \pi^{\mathcal{Q}}(H)=\sup_{Q \in \mathcal{Q}} \pi^{Q}(H)=\sup_{P \in \mathcal{P}} \pi^{P}(H)=\pi^{\mathcal{P}}(H)=\underline{\pi}^{{\mathcal{Q}}}(H).
\end{align}
\end{small}
\end{proposition}
Note that   \cite[Proposition 21]{Burz20} shows in a one step market that $\pi^{\mathcal{Q}}(H)=\sup_{Q \in \mathcal{Q}} \pi^{Q}(H).$ 
%\red{ (NOT SURE IF WE NEED TO ADD THE REST) which is similar to the third equality since for $P \in  \mathcal{Q}$ such that $NA(P)$ does not hold true, $\p^{P}(H)=-\infty$.}
\begin{proof}
Assumption \ref{PaQ}  $iii)$  implies that for all $Q \in  \mathcal{Q}$, there exists $P \in  \mathcal{P}$, such that $\pi^{Q}(H) \leq \pi^{P}(H)$. 
%Indeed, this follows from the fact that  $\{x \in \mathbb{R}, \; \exists \phi \in \Phi, \; V_{T}^{x,\phi} \geq H\ {P} \mbox{-a.s.}\} \subset \{x \in \mathbb{R}, \; \exists \phi \in \Phi, \; V_{T}^{x,\phi} \geq H\ {Q} \mbox{-a.s.}\}.$ 
So, $\sup_{Q \in \mathcal{Q}} \pi^{Q}(H)  \leq \sup_{P \in \mathcal{P}} \pi^{P}(H).$ 
As the set $\overline{\mathcal{P}}$ of Theorem \ref{spa} is such that $\overline{\mathcal{P}} \subset \mathcal{Q},$  we get that 
$$\sup_{Q \in \mathcal{Q}} \pi^{Q}(H)  =\sup_{ P  \in \overline{\mathcal{P}}} \pi^{P}(H).$$
Now, Theorem \ref{super} shows that $\pi^{\mathcal{Q}}(H)=\sup_{P \in \mathcal{P}} \pi^{P}(H)=\pi^{\mathcal{P}}(H)$ and Theorem \ref{spa} that $\sup_{ P  \in \overline{\mathcal{P}}} \pi^{P}(H)=\pi^{\mathcal{Q}}(H),$ which proves the first three equalities. 
Finally, recalling the proof of  the first equality in  \eqref{tordtord}, we get that   $\underline{\pi}^{{\mathcal{Q}}}(H) = \sup_{Q \in \mathcal{Q}} \pi^{Q}(H).$  %we use the same arguments  as in the first equality of  \eqref{pricephi}. Let $\varepsilon>0$ and $Q \in \mathcal{Q}$, then $V_{T}^{\underline{\pi}^{{\mathcal{Q}}}(H)+\varepsilon} \geq H$ $Q$-as, so   $\underline{\pi}^{{\mathcal{Q}}}(H)+\varepsilon \geq  \pi^{Q}(H)$.   As this is true for all  $\varepsilon>0$ and $Q \in \mathcal{Q}$, we get that $\underline{\pi}^{{\mathcal{Q}}}(H) \geq \sup_{Q \in \mathcal{Q}} \pi^{Q}(H)$.  The equality follows since $\underline{\pi}^{{\mathcal{Q}}}(H) \leq  \pi^{\mathcal{Q}}(H)$.

\end{proof}
We want now  to find a set $\mathcal{P}$ such that the probability measure $\widehat P \in \mathcal{P}$  have the same  support as $\mathcal{Q}$. 
To do that, we first recall the definition of the supports. Then, we recall in Lemma \ref{supportcar} some properties of those supports. Lemmata \ref{lemmptilde} and \ref{lemmaoa} allow to prove in Proposition \ref{PPstar} the existence of some measure $\widetilde{P}$ having the same support as $\mathcal{Q}$ and such that  the local NA($\widetilde{P}$) holds true 
$\mathcal{Q}^t$-q.s. for all $t.$ From $\widetilde{P},$ we construct by recursion a set of priors $\widetilde{\mathcal{P}}$ satisfying Assumption \ref{PaQ}, see Proposition \ref{Att}. The application of Theorem \ref{super} is given in Theorem \ref{theo2}. 

% \footnote{ Note that in \citep[Theorem 3.6]{BC19} the support of the measures in $\mathcal{P}$ generates the same affine hull as the  support of $\mathcal{Q}$ but the support do not necessarily coicinde.}\\
Let $P \in \mathfrak{P}\left(\Omega^{T}\right)$ with the fixed disintegration $P:=P_1 \otimes p_{2} \otimes \cdots \otimes p_{T}$ where $P_1 \in \mathfrak{P}(\O_{1})$ and $p_{t} \in \mathcal{SK}_{t}$ for all $2 \leq t \leq T$. 
For all  $0 \leq t \leq T-1$,  the random sets 
%${E}^{t+1}: \; \Omega^{t} \times \mathfrak{P}(\O_{t+1}) \twoheadrightarrow \mathbb{R}^{d}$, 
${D}_{\mathcal{Q}}^{t+1},\ {D}_{P}^{t+1} \; : \Omega^{t} \twoheadrightarrow \mathbb{R}^{d}$  are defined  by
\begin{align}
%\label{DefE}
%{E}^{t+1}(\o^{t},p)&:=\bigcap  \left\{ A \subset \mathbb{R}^{d},\; \mbox{closed}, \; p\left(\Delta S_{t+1}(\o^{t},.) \in A\right) =1\right\} \mbox{for } p \in \mathfrak{P}(\O_{t+1}),\\
\label{dedD}
{D}_{\mathcal{Q}}^{t+1}(\o^{t})&:=\bigcap  \left\{ A \subset \mathbb{R}^{d},\; \mbox{closed}, \; q\left(\Delta S_{t+1}(\o^{t},.) \in A\right)=1, \; \forall \,q \in \mathcal{Q}_{t+1}(\o^{t}) \right\},\\
\label{DefPDARB1}
{D}_{P}^{t+1}(\o^{t})&:=\bigcap  \left\{ A \subset \mathbb{R}^{d},\; \mbox{closed}, \; p_{t+1}\left(\Delta S_{t+1}(\o^{t},.) \in A,\o^{t}\right) =1\right\}.
\end{align}
The following  lemma is proved in \citep[Theorem 12.14]{Hitch}, \citep[Lemma 4.2]{BN} and \citep[Lemma 5.2]{BC19}. 
\begin{lemma}
\label{supportcar}
 Let $0 \leq t \leq T-1$, $\o^t \in \O^t$ be fixed and ${P}\in \mathfrak{P}(\O^{T})$ with the fixed disintegration ${P} :={P}_{1} \otimes {p}_{2} \otimes  \cdots \otimes {p}_{T}.$
 \begin{itemize}
\item [i)] Then, $p_{t+1}\left(\Delta S_{t+1}(\o^{t},.) \in {D}_{P}^{t+1}(\o^{t}),\o^t\right)=1$ and ${D}_{P}^{t+1}(\o^{t})$ is the smallest closed set $A \subset \mathbb{R}^{d}$ such that $p_{t+1}\left(\Delta S_{t+1}(\o^{t},.) \in A,\o^{t}\right) =1.$ 
Moreover,  $h \in {D}_{P}^{t+1}(\o^{t})$ if and only if $p_{t+1}\left(\Delta S_{t+1}(\o^{t},.) \in B\left(h,{1}/{k}\right),\o^t\right)>0$ for all $k \geq 1$. 
 \item [ii)] For all $q \in \mathcal{Q}_{t+1}(\o^{t})$,  $q\left( \Delta S_{t+1}(\o^{t},.) \in {D}_{\mathcal{Q}}^{t+1}(\o^{t})\right)=1$ and
 ${D}_{\mathcal{Q}}^{t+1}(\o^{t})$ is the smallest closed set $A \subset \mathbb{R}^{d}$ such that $q\left(\Delta S_{t+1}(\o^{t},.) \in A\right) =1$ for all  $q \in \mathcal{Q}_{t+1}(\o^{t})$. 
 Moreover, 
 $h \in {D}_{\mathcal{Q}}^{t+1}(\o^{t})$ if and only if for all $k \geq 1$, there exists  $q^{k} \in \mathcal{Q}_{t+1}(\o^{t})$ such that  $q^{k}\left( \Delta S_{t+1}(\o^{t}) \in B\left(h,{1}/{k}\right)\right)>0$.
 \end{itemize}
\end{lemma}

\begin{lemma}
\label{lemmptilde}
Assume that Assumptions  \ref{SassARB} and \ref{QanalyticARB} hold true. Then, there exists 
some $\widetilde{P}\in \mathfrak{P}(\O^{T})$ 
such that   $D_{\widetilde{P}}^{t+1}(\cdot)=D_{\mathcal{Q}}^{t+1}(\cdot)$ \footnote{The equality holds true for all $\o^{t}$}  for all $0 \leq t \leq T-1.$
\end{lemma}
\begin{proof}
Fix some  $0 \leq t \leq T-1$. The random set $\mathcal{D}_{\mathcal{Q}}^{t+1}$ is closed valued and  $\mathcal{B}_{c}(\O^{t})$-measurable, see \citep[Lemma 4.3]{BN}). Hence, $\mathcal{D}_{\mathcal{Q}}^{t+1}$ admits a Castaing representation and 
there exists a sequence of $\mathcal{B}_{c}(\O^{t})$-measurable random variables  $(X_n)_{n \geq 1}$ such that  for all $\o^{t} \in \O^{t}$,  $\overline{\{X_{n}(\o^{t}), n \geq 1\}}= \mathcal{D}_{\mathcal{Q}}^{t+1}(\o^{t})$, see for example \citep[Theorem 14.5]{rw}. 
For some $n,k \geq 1$, let 
\begin{align}
\label{Pnk}
\Psi_{t+1}^{n,k}(\o^{t}):=\left\{p \in \mathcal{Q}_{t+1}(\o^{t}), p\left( \Delta S_{t+1}(\o^{t},\cdot) \in B\left(X_{n}(\o^{t}), \frac{1}{k}\right)\right)>0\right\}.
\end{align}
Let $\o^t \in \O^t$. Then, $X_{n}(\o^{t}) \in D_{\mathcal{Q}}^{t+1}(\o^{t})$ and Lemma \ref{supportcar} shows that for all $k\geq 1,$ 
$\exists p \in \mathcal{Q}_{t+1}(\o^{t}),$ such that  $p\left( \Delta S_{t+1}(\o^{t},\cdot) \in B\left(X_{n}(\o^{t}), {1}/{k}\right)\right)>0$, i.e., $p \in \Psi_{t+1}^{n,k}(\o^{t})$. Thus,  $\O^{t}=\{\Psi_{t+1}^{n,k} \neq \emptyset\}$.  
As $\o^{t} \mapsto X_{n}(\o^{t})$ is $\mathcal{B}_{c}(\O^{t})$-measurable, $(\o^{t},\o_{t+1}) \mapsto 1_{\Delta S_{t+1}(\cdot,\cdot) \in B\left(X_{n}(\cdot),{1}/{k}\right)}(\o^{t},\o_{t+1}) $ is $\mathcal{B}_{c}(\O^{t})\otimes \mathcal{B}(\O_{t+1})$-measurable. We get that 
$(\o^{t},p) \mapsto p\left( \Delta S_{t+1}(\o^{t},\cdot) \in B\left(X_{n}(\o^{t}), {1}/{k}\right)\right)$ is  $\mathcal{B}_{c}(\O^{t})\otimes \mathcal{B}\left(\mathfrak{P}(\O_{t+1})\right)$-measurable by a monotone class argument.
Let for some Polish space $X$ and  some paving $\mathcal{J}$ (i.e.,  a non-empty collection of subsets of $X$ containing the empty set),  $\mathfrak{A}(\mathcal{J})$ denotes the set of all nuclei of Suslin Scheme on $\mathcal{J}$ (see \citep[Definition 7.15 p157]{bs}). 
Then, 
 $$\mbox{graph}\left(\Psi_{t+1}^{n,k}\right) \in  \mathfrak{A}\left(\mathcal{B}_{c}(\O^{t})\otimes \mathcal{B}\left(\mathfrak{P}(\O_{t+1})\right)\right),$$ 
using \citep[Proposition 7.35 p158]{bs} together with Assumption \ref{QanalyticARB}.
Thus,  \citep[Lemma 4.11]{BN} (which relies on \citep{Lee78})  gives the existence of $p_{t+1}^{n,k}(\cdot,\o^{t}) \in   \mathcal{S}K_{t+1}$ such that for all  $\o^{t} \in \O^{t}=\{\Psi_{t+1}^{n,k} \neq \emptyset\}=\mbox{proj}_{\O^t}\mbox{graph}\left(\Psi_{t+1}^{n,k}\right)$, $p_{t+1}^{n,k}(\cdot,\o^{t})  \in \Psi_{t+1}^{n,k}(\o^t)$. 
We set \begin{align}
\label{PstarM}
\widetilde{p}_{t+1}(\cdot,\o^{t}) & := \sum_{n \geq 1} \sum_{k \geq 1} \frac{1}{2^{n+k}} p_{t+1}^{n,k}(\cdot,\o^{t}) \in \mathfrak{P}({\Omega}_{t+1}) \\
 \label{Ptilde}
 \widetilde{P} & := \widetilde{p}_{1} \otimes \cdots \otimes \widetilde{p}_{T}.
 \end{align}
 We prove now that  $D_{\widetilde{P}}^{t+1}(\o^{t})=D_{\mathcal{Q}}^{t+1}(\o^{t})$ for $\o^{t} \in   \O^{t}$. 
% To do so, we prove  that ${E}^{t+1}(\o^{t},\widetilde{p}_{t+1}(\cdot,\o^{t}))=D_{\mathcal{Q}}^{t+1}(\o^{t})$ for $\o^{t} \in   \O^{t}$.  
Fix some  $\o^{t} \in   \O^{t}$. As for all $n,k \geq 1$, $ p_{t+1}^{n,k}(\cdot,\o^{t}) \in \mathcal{Q}_{t+1}(\o^{t}),$ we have that  
 $$\widetilde{p}_{t+1}(\Delta S_{t+1}(\o^{t},\cdot) \in D_{\mathcal{Q}}^{t+1}(\o^{t}),\o^t)= \sum_{n \geq 1} \sum_{k \geq 1} \frac{1}{2^{n+k}} p_{t+1}^{n,k} (\Delta S_{t+1}(\o^{t},\cdot) \in D_{\mathcal{Q}}^{t+1}(\o^{t}),\o^t)=1.$$  
 Thus, $D_{\widetilde{P}}^{t+1}(\o^{t}) \subset D_{\mathcal{Q}}^{t+1}(\o^{t}).$
%   and ${E}^{t+1}(\o^{t},\widetilde{p}_{t+1}(\cdot,\o^{t})) \subset D_{\mathcal{Q}}^{t+1}(\o^{t})$
 Now, fix some  $n \geq1.$ For all $k \geq 1$, 
\begin{small}
 \begin{align*}
 \widetilde{p}_{t+1}\left( \Delta S_{t+1}(\o^{t},\cdot) \in B\left(X_{n}(\o^t),\frac{1}{k}\right),\o^{t}\right) \geq \frac{1}{2^{n+k}} \; p_{t+1}^{n,k}\left(\Delta S_{t+1}(\o^{t},\cdot) \in  B\left(X_{n}(\o^t),\frac{1}{k}\right),\o^{t}\right)>0
 \end{align*}
 \end{small} and using  Lemma \ref{supportcar}, 
  $X_{n}(\o^t)  \in D_{\widetilde{P}}^{t+1}(\o^{t})$. 
  So, $D_{\mathcal{Q}}^{t+1}(\o^{t})=\overline{\{X_{n}(\o^{t}), n \geq 1\}} \subset  D_{\widetilde{P}}^{t+1}(\o^{t})$. 
% $X_{n}(\o^t)  \in {E}^{t+1}(\o^{t},\widetilde{p}_{t+1}(\cdot,\o^{t}))$.  
 %So, $D_{\mathcal{Q}}^{t+1}(\o^{t})=\overline{\{X_{n}(\o^{t}), n \geq 1\}} \subset  {E}^{t+1}(\o^{t},\widetilde{p}_{t+1}(\cdot,\o^{t}))$. \\
\end{proof}

Let  $0 \leq t \leq T-1$ and $\o^t \in \O^t$. 
We say that the local quasi-sure no-arbitrage condition NA($\mathcal{Q}_{t+1})(\o^{t})$ holds true if 
$h\Delta S_{t+1}(\o^{t},\cdot) \geq 0 \; \mathcal{Q}_{t+1}(\o^{t}) \mbox{-q.s.}$ for some $h \in \mathbb{R}^{d}$ implies that $h\Delta S_{t+1}(\o^{t},\cdot) = 0 \;\mathcal{Q}_{t+1}(\o^t)\mbox{-q.s.}$ 
We denote by 
$$ N^{t}:=\{\o^{t} \in \O^{t},\; \mbox{NA($\mathcal{Q}_{t+1})(\o^{t})$ fails}\}.$$ 
Moreover, let ${P}\in \mathfrak{P}(\O^{T})$ with the fix disintegration ${P} :={P}_{1} \otimes {p}_{2} \otimes  \cdots \otimes {p}_{T}.$  Similarly, the local  no-arbitrage condition  
NA$({P})(\o^{t})$ holds true if $h \Delta S_{t+1}(\o^t,\cdot) \geq 0$ ${p}_{t+1}(\cdot,\o^t)$-a.s. for some $h \in \mathbb{R}^d$ implies that $h \Delta S_{t+1}(\o^t,\cdot) = 0$ ${p}_{t+1}(\cdot,\o^t)$-a.s.

\begin{lemma}
\label{lemmaoa}
Assume that there exists 
some $\widetilde{P}\in \mathfrak{P}(\O^{T})$ 
such that   $D_{\widetilde{P}}^{t+1}(\cdot)=D_{\mathcal{Q}}^{t+1}(\cdot),$  for all $0 \leq t \leq T-1.$ 
Then, $$ N^{t}=\{\o^{t} \in \O^{t},\; \mbox{NA($\widetilde{P})(\o^{t})$ fails}\}.$$ 
\end{lemma}
\begin{proof}
Consider the fix disintegration $\widetilde{P} :=\widetilde{P}_{1} \otimes \widetilde{p}_{2} \otimes  \cdots \otimes \widetilde{p}_{T}.$ 
Let $0 \leq t \leq T-1$.   Let $\o^{t} \in  \O^{t},$  $h \in \mathbb{R}^{d}.$ Then, 
\begin{align}
\label{pareil}
h \Delta S_{t+1}(\o^{t}, \cdot) \geq  0  \; \mbox{${Q}_{t+1}(\o^{t})$-q.s.} & \Longleftrightarrow  h y  \geq 0 ,\;  \forall y \in  D_{\mathcal{Q}}^{t+1}(\o^{t})=D_{\widetilde{P} }^{t+1}(\o^{t}) \\
& \Longleftrightarrow  h\Delta S_{t+1}(\o^{t}, \cdot) \geq 0 \; \mbox{$\widetilde{p}_{t+1}(\cdot,\o^{t})$-a.s.}
\end{align}
We prove only the first equivalence as the second is similar. The reverse implication  follows from 
$\{\Delta S_{t+1}(\o^{t}, \cdot) \in D_{\mathcal{Q}}^{t+1}(\o^{t})\} \subset \{ h\Delta S_{t+1}(\o^{t}, \cdot) \geq 0 \}.$
We prove the contraposition of the direct implication. Assume that there exists some $y_0 \in   D_{\mathcal{Q}}^{t+1}(\o^{t})$, such that $h y_0 < 0$. Then, there exists some $\varepsilon>0$ such that for all $y \in B(y_0,\varepsilon),$ the open ball of center $y_0$ and radius $\e>0$, $h y < 0$. Now, as $y_0 \in D_{\mathcal{Q}}^{t+1}(\o^{t}),$ Lemma \ref{supportcar} shows that there exists also some $q \in \mathcal{Q}_{t+1}(\o^{t})$ such that $q(\Delta S_{t+1}(\o^{t}, \cdot) \in  B(y_0,\varepsilon))>0$. As $\{\Delta S_{t+1}(\o^{t}, \cdot)  \in B(y_0,\varepsilon)\} \subset \{h\Delta S_{t+1}(\o^{t}, \cdot)  < 0 \}$, we get that $q( h\Delta S_{t+1}(\o^{t}, \cdot)  < 0)>0$. Using similar arguments, we get that 
\begin{align}
\label{pareil2}
\exists q \in  \mathcal{Q}_{t+1}(\o^{t}),\;  q(h \Delta S_{t+1}(\o^{t}, \cdot)>0) >  0  & \Longleftrightarrow   \exists y_0 \in  D_{\mathcal{Q}}^{t+1}(\o^{t})=D_{\widetilde{P} }^{t+1}(\o^{t}),\, \;  h y_0 >0 \\
&\Longleftrightarrow \widetilde{p}_{t+1}(h\Delta S_{t+1}(\o^{t}, \cdot)>0,\o^t)>  0.  
\end{align}
%So, $\o^{t} \in  \{\o^{t} \in \O^{t},\; \mbox{NA$(\widetilde P)(\o^{t})$ fails}\}$. The proof of the reverse implication is similar.
\end{proof}

In the following proposition, we prove the existence of a probability measure $\widetilde{P}\in \mathfrak{P}(\O^{T})$ having the same support as $\mathcal{Q}$ and  such that
NA$(\widetilde{P})(\cdot)$ holds true $\mathcal{Q}$-q.s. The main difference with the probability measure $P^*$ given by \citep[Theorem 3.29]{BC19} is that 
$\widetilde{P}$ does not belong to $\mathcal{P}.$ However, only the affine hull generated by the  the support of $P^*$ coincides with the one generated by the support of $\mathcal{Q}.$% We want now  to find a set $\mathcal{P}$ such that the probability measures $\widehat P \in \mathcal{P}$  have the same  support as $\mathcal{Q}$ \footnote{ Note that in \citep[Theorem 3.6]{BC19} the support of the measures in $\mathcal{P}$ generates the same affine hull as the  support of $\mathcal{Q}$ but the support do not necessarily coicinde.}\\
\begin{proposition}
\label{PPstar}
Assume that Assumptions  \ref{SassARB} and \ref{QanalyticARB} hold true. The $NA(\mathcal{Q})$ condition holds true if and only if there exists 
some $\widetilde{P}\in \mathfrak{P}(\O^{T})$ with the fix disintegration $\widetilde{P} :=\widetilde{P}_{1} \otimes \widetilde{p}_{2} \otimes  \cdots \otimes \widetilde{p}_{T}$ 
such that for all $0 \leq t \leq T-1$ \begin{enumerate}
 \item[i)]    $D_{\widetilde{P}}^{t+1}(\cdot)=D_{\mathcal{Q}}^{t+1}(\cdot)$
 \item[ii)]    The set  $\{\o^{t} \in \O^{t},\; \mbox{NA$(\widetilde{P})(\o^{t})$ fails}\}$ is  a $\mathcal{Q}^{t}$-polar set.
 \end{enumerate}
\end{proposition}
\begin{proof}
{\it Reverse implication.} 
For all $0 \leq t \leq T-1$, 
i) and $ii)$ together with Lemma \ref{lemmaoa}, show that $N^t$ is  a $\mathcal{Q}^{t}$-polar set. Thus, the $NA(\mathcal{Q})$ condition holds true using  \citep[Theorem 4.5]{BN}).\\
{\it Direct implication.} 
Fix some  $0 \leq t \leq T-1$. Lemmata  \ref{lemmptilde} and \ref{lemmaoa} together with \citep[Theorem 4.5]{BN}) shows that $i)$ and $ii)$ hold true. \end{proof}

Let $\widetilde{P}$ as in Proposition \ref{PPstar}. For all $1\leq t \leq T-1$, let $\widetilde{\mathcal{P}}_{t+1}: \Omega^t \twoheadrightarrow \mathfrak{P}(\O_{t+1})$ be defined for all $\o^{t} \in \O^{t}$ by
  \begin{align}
 \label{splitbis}
\widetilde{\mathcal{P}}_{t+1}(\o^{t}):=
\left\{ \l \widetilde{p}_{t+1}(\cdot,\o^{t})+ (1-\l) q, \;q \in \mathcal{Q}_{t+1}(\o^{t}),\;  0<\l\leq 1 \right\}.
\end{align}
%where $\widetilde{p}_{t+1}$  comes from  Lemma  \ref{lemmptilde}. 
Let
$\widetilde{\mathcal{P}}:=\widetilde{\mathcal{P}}^T$ be defined recursively as follows: $\widetilde{\mathcal{P}}^{1}:=\left\{\l \widetilde P_{1}+ (1-\l)P,\; P \in \mathcal{Q}_{1},\; 0<\l\leq 1  \right\}$
and for all $1 \leq t \leq T-1$
  \begin{align}
\label{PstarARB}
\widetilde{\mathcal{P}}^{t+1}&:=\Bigl\{P \otimes  q, \; P \in \widetilde{\mathcal{P}}^{t},\, q \in \mathcal{SK}_{t+1}, \; q(\cdot,\o^{t}) \in \widetilde{\mathcal{P}}_{t+1}(\o^{t}) \; \mbox{for all $\o^{t} \in \O^{t}$}  \Bigr\}.
\end{align}
\begin{proposition}
\label{Att}
Let Assumptions \ref{SassARB}, \ref{QanalyticARB} and  the $NA(\mathcal{Q})$ condition hold true. 
Then, Assumption \ref{PaQ} holds true for $\widetilde{\mathcal{P}}$ defined above. Moreover, for all 
$P \in \widetilde{\mathcal{P}}$ and  $0\leq t \leq T-1,$ 
  \begin{align}
\label{moon}
D_{P}^{t+1}(\cdot)=D_{\mathcal{Q}}^{t+1}(\cdot)=D_{\widetilde{\mathcal{P}}}^{t+1}(\cdot)
\end{align}
\end{proposition}
\begin{proof}
%So, there exists $(\l_{t+1})_{0\leq t \leq T-1} \subset (0,1],$ $Q^1 \in \Qc^1$ and 
%$(q_{t+1})_{1\leq t \leq T-1}$ such that $q_{t+1}(\cdot,\o^{t})  \in \mathcal{Q}_{t+1}(\o^{t})$ and 
%$p_{t+1}(\cdot,\o^{t})=\l_{t+1} \widetilde{p}_{t+1}(\cdot,\o^{t})+ (1-\l_{t+1}) q_{t+1}(\cdot,\o^{t}).$ 
%Theorem \ref{PPstar} shows that   $D_{\widetilde{P}}^{t+1}(\cdot)=D_{\mathcal{Q}}^{t+1}(\cdot)$ and thus 
%$$p_{t+1}\left(D_{\mathcal{Q}}^{t+1}(\o^{t}),\o^{t}\right)=\l_{t+1} \widetilde{p}_{t+1}\left(D_{\widetilde{P}}^{t+1}(\o^{t}),\o^{t}\right)+ (1-\l_{t+1}) q_{t+1}\left(D_{\mathcal{Q}}^{t+1}(\o^{t}),\o^{t}\right)=1.$$ 
%So, $D_{P}^{t+1}(\cdot) \subset D_{\mathcal{Q}}^{t+1}(\cdot)$. 
%Let ${Q}\in \mathcal{Q}$ with the fix disintegration $Q :=q_{1} \otimes   \cdots \otimes q_{T}$  where $q_{t+1}(\cdot,\o^{t})  \in \mathcal{Q}_{t+1}(\o^{t}).$  
%Then, $\frac1n \widetilde{p}_{t+1}(\cdot,\o^{t})+ \left(1-\frac1n \right) q_{t+1}(\cdot,\o^{t}) \in \widetilde{\mathcal{P}}_{t+1}(\o^{t})$ and 
% $$1=\frac1n \widetilde{p}_{t+1}\left(D_{\widetilde{\mathcal{P}}}^{t+1}(\o^{t}),\o^{t}\right)+ \left(1-\frac1n \right) q_{t+1}\left(D_{\widetilde{\mathcal{P}}}^{t+1}(\o^{t}),\o^{t}\right)=\frac1n + \left(1-\frac1n \right) q_{t+1}\left(D_{\widetilde{\mathcal{P}}}^{t+1}(\o^{t}),\o^{t}\right).$$
%Letting $n$ go to infinity, we get that  $q_{t+1}\left(D_{\widetilde{\mathcal{P}}}^{t+1}(\o^{t}),\o^{t}\right)=1$ and thus 
%$D_{\mathcal{Q}}^{t+1}(\cdot) \subset D_{\widetilde{\mathcal{P}}}^{t+1}(\o^{t})$
{\it Proof of Assumption \ref{PaQ} iii)} Exactly as in \citep[Lemma 3.9]{BC19}, for $Q \in \mathcal{Q}$  with the fix  disintegration $Q:=Q_{1}\otimes q_{2} \otimes \cdots \otimes q_{T},$  
 there exists some $(\widetilde{R}_{k})_{0 \leq k \leq T-1} \subset \mathfrak{P}(\O^{T})$ such that
%for all $1 \leq t \leq T,$
% there exist some $(\widetilde{R}_{k}^{t})_{0 \leq k \leq t-1} \subset \mathfrak{P}(\O^{t})$ such that   for all $n>1$,
%\begin{align}
%\label{densityt}
% P^t_{n}:=\left(1-\frac{1}{n}\right)^{t} Q^t + \frac1{n^t}\sum_{k=0}^{t-1} (n-1)^k \widetilde{R}_{k}^{t} \in  \widetilde{\mathcal{P}}^{t}.  \end{align}
%Thus, for some $Q \in \mathcal{Q}$
$${P}:=\frac{1}{2^T}Q+ \frac1{2^T}\sum_{k=0}^{T-1} \widetilde{R}_{k} \in  \widetilde{\mathcal{P}}.$$
It is clear that $Q \ll  P$ and $iii)$ in Assumption \ref{PaQ}  is proved. \\
{\it Proof of Assumption \ref{PaQ} ii)} 
We show that for all $ 0\leq t \leq T$, $\mathcal{Q}^{t}$-polar sets  are also $\widetilde{\mathcal{P}}^{t}$-polar sets.  Note that this implies that $\mathcal{Q}^{t}$-polar sets  are also $\widetilde{{P}}^{t}$-null sets. \\
From \eqref{PstarM}, it is clear that  for all $0\leq t \leq T$, all $\o^{t} \in \O^{t},$ 
\begin{align}
\label{usefull}
\mbox{$\mathcal{Q}_{t+1}(\o^{t})$-polar sets are also  $\widetilde{p}_{t+1}(\cdot,\o^{t})$-null sets.}
\end{align} 
 We prove  by induction that any $\mathcal{Q}^{t}$-polar set is a ${P}^{t}$-null set for all $P^t \in \widetilde{\mathcal{P}}^{t}$. For $t=1$, this follows from \eqref{usefull} and  $\widetilde{\mathcal{P}}^{1}=\left\{\l \widetilde P_{1}+ (1-\l)P,\; P \in \mathcal{Q}_{1},\; 0<\l\leq 1  \right\}.$

 We assume now that this is the case for $1 \leq t \leq T-1$. Let $N$ be a $\mathcal{Q}^{t+1}$ polar set and let $A \in \mathcal{B}(\O^{t+1})$ such that $N \subset A$ and $Q(A)=0$ for all $Q \in \mathcal{Q}^{t+1}.$ Let $P^{t+1}= P^t \otimes  p_{t+1} \in \widetilde{\mathcal{P}}^{t+1}$, where  $P^t \in \widetilde{\mathcal{P}}^{t}$ and  $ p_{t+1}(\cdot,\o^{t}) \in \widetilde{\mathcal{P}}_{t+1}(\o^{t}),$ for all $\o^{t} \in \O^{t}.$
 
Let $A_{\o^{t}}:= \{\o_{t+1} \in \O_{t+1},\; (\o^t,\o_{t+1}) \in A \} \in \mathcal{B}(\O_{t+1})$ be the section of $A$ along $\o^t,$ see \cite[Proposition 4.46]{Hitch}. 
First, we prove that 
\begin{align}
\label{At}
B:=\{\o^{t} \in \O^t,\; \exists q \in \mathcal{Q}_{t+1}(\o^{t}),\; q(A_{\o^{t}})>0\} \in \mathcal{B}_{c}(\O^{t})
\end{align}
and that $B$  is a  $\mathcal{Q}^{t}$-polar set. 
We have that,
$$B=\mbox{proj}_{\O^{t}}\{(\o^{t}, q), \; q \in \mathcal{Q}_{t+1}(\o^{t}),\; q(A_{\o^{t}})>0\}=\mbox{proj}_{\O^{t}} \left[ \mbox{graph}\left(\mathcal{Q}_{t+1}\right) \cap C\right],$$
where $C:=\{(\o^{t}, q) \in \O^{t} \times  \mathfrak{P}(\O_{t+1}), q(A_{\o^{t}})>0\} \in \mathcal{B}(\O^{t})\otimes \mathcal{B}(\frak{P}(\O_{t+1})),$ as 
$$\O^{t} \times  \mathfrak{P}(\O_{t+1}) \ni (\o^{t},q) \mapsto q(A_{\o^{t}})=\int_{\O_{t+1}} 1_{A}(\o^{t},\o_{t+1})q(d\o_{t+1})$$ is $\mathcal{B}(\O^{t})\otimes \mathcal{B}(\frak{P}(\O_{t+1}))$ measurable by a monotone class argument. 
So,  recalling Assumption \ref{QanalyticARB},  \cite[Corollary 7.35.2 p160, Proposition 7.36 p161, Proposition 7.39 p165]{bs}), $\mbox{graph}\left(\mathcal{Q}_{t+1}\right) \cap C$ and $B$ are analytic sets. 
Furthermore, the Jankov-von Neumann Theorem (see  \cite[Proposition 7.49 p182]{bs}) implies that there exists $\widehat q_{t+1}   \in \mathcal{S}K_{t+1}$ such that 
for all $\o^{t} \in B$, $\widehat q_{t+1}(\cdot,\o^{t}) \in \mathcal{Q}_{t+1}(\o^{t})$ and $\widehat q_{t+1}(A_{\o^{t}},\o^{t})>0$. \\
For $\o^t \in \O^t \setminus B,$ $\widehat q_{t+1}(\cdot,\o^{t})$ is chosen equal to some given kernel in $\mathcal{Q}_{t+1}(\o^{t})$. 
Now, assume that there exists some $\widehat Q^{t} \in \mathcal{Q}^{t}$, such that $\widehat Q^{t}(B)>0$. Let $\widehat Q^{t+1}:=\widehat Q^{t} \otimes \widehat q_{t+1} \in \mathcal{Q}^{t+1}$. We have that
$$\widehat Q^{t+1}(A) \geq \int_{B} \widehat q_{t+1}(A_{\o^{t}},\o^{t})\widehat Q^{t}(d\o^{t})>0,$$
which contradicts that   $\widehat Q^{t+1}(A)=0.$ Thus,  $B$ is $\mathcal{Q}^{t}$-polar set and by the induction hypothesis,  $ P^{t}(B)=0$ also. Fix some $\o^{t} \notin B$: for all $q \in \mathcal{Q}_{t+1}(\o^{t})$ , $ q(A_{\o^{t}},\o^{t})=0$. 
Hence, $A_{\o^{t}}$ is a $\mathcal{Q}_{t+1}(\o^{t})$-polar set and, recalling \eqref{usefull}, also  a $\widetilde{p}_{t+1}(\cdot,\o^{t})$-null set. 
 By definition of $\widetilde{\mathcal{P}}_{t+1}(\o^{t})$, we also get that $ p_{t+1}(A_{\o^{t}},\o^t)=0$. 
Finally, we have that
\begin{align}
\label{PtQT}
{P}^{t+1}(A)=\int_{B} {p}_{t+1}(A_{\o^{t}},\o^{t}) {P}^{t}(d\o^{t})+\int_{\O^{t} \backslash{B}} {p}_{t+1}(A_{\o^{t}},\o^{t}) {P}^{t}(d\o^{t})=0.
\end{align}
Thus, $N$ is also a ${P}^{t+1}$-null set, which concludes the proof of $ii)$. 

{\it Proof of \eqref{moon}} 
As Assumption \ref{PaQ} $ii)$ and $iii)$ hold true, Remark \ref{lemem} shows that $\widetilde{\mathcal{P}}$ and $\mathcal{Q}$ have the same polar sets. So, Lemma \ref{supportcar} shows that for all $0\leq t \leq T-1$,
$D_{\mathcal{Q}}^{t+1} (\cdot) = D_{\widetilde{\mathcal{P}}}^{t+1}(\cdot).$
Let $P\in \widetilde{\mathcal{P}}.$ Then, Lemma \ref{supportcar} shows that  $D_{P}^{t+1}(\cdot) \subset D_{\widetilde{\mathcal{P}}}^{t+1}(\cdot)$ for all $0\leq t \leq T-1.$
Set $P=P_1\otimes  p_2 \otimes \ldots \otimes p_T $ where $P_1 \in \widetilde{\mathcal{P}}^{1}$ and $p_{t+1}(\cdot,\o^{t}) \in \widetilde{\mathcal{P}}_{t+1}(\o^{t})$ for all $\o^{t} \in \O^{t}$  and for all $1\leq t \leq T-1$. 
Fix  $0\leq t \leq T-1$ and $\o^{t} \in \O^{t}$.  There exists $\l_{t+1}\in (0,1]$ and 
 $q_{t+1}  \in \mathcal{Q}_{t+1}(\o^{t})$ such that
$p_{t+1}(\cdot,\o^{t})=\l_{t+1} \widetilde{p}_{t+1}(\cdot,\o^{t})+ (1-\l_{t+1}) q_{t+1}(\cdot).$ 
Then, Lemma \ref{supportcar} shows that
$$1=p_{t+1}\left(D_{P}^{t+1}(\o^{t}),\o^{t}\right)=\l_{t+1} \widetilde{p}_{t+1}\left(D_{P}^{t+1}(\o^{t}),\o^{t}\right)+ (1-\l_{t+1}) q_{t+1}\left(D_{P}^{t+1}(\o^{t})\right).$$ 
As $\l_{t+1}>0,$ $\widetilde{p}_{t+1}\left(D_{P}^{t+1}(\o^{t}),\o^{t}\right)=1$ and using Proposition \ref{PPstar} and  Lemma \ref{supportcar}, we get that 
$D_{\widetilde{\mathcal{P}}}^{t+1}(\o^{t})=D_{\mathcal{Q}}^{t+1} (\o^{t})=
D_{\widetilde P}^{t+1}(\o^{t}) \subset D_{P}^{t+1}(\o^{t})$. As this is true for all $\o^{t} \in \O{^t}$, 
 \eqref{moon} is proved. 

{\it Proof of Assumption \ref{PaQ} i)} Let $P \in \widetilde{\mathcal{P}}$. Fix $0\leq t \leq T-1,$ \eqref{moon} shows that  $D_{{P}}^{t+1}(\cdot)=D_{\mathcal{Q}}^{t+1}(\cdot).$ Lemma \ref{lemmaoa},  $NA(\mathcal{Q})$ condition and  \citep[Theorem 4.5]{BN}) show that $\{\o^{t} \in \O^{t},\; \mbox{NA(${P})(\o^{t})$ fails}\}$ is 
 $\mathcal{Q}^{t}$-polar. Using $ii)$, it is also a 
${P}^{t}$-null set. The usual equivalence between local no-arbitrage and global no-arbitrage shows that NA$({P})$ holds  true: $i)$ follows.  
\end{proof}

We are now in position to state our second application.

\begin{theorem}
\label{theo2}
Assume that Assumptions  \ref{SassARB},  \ref{QanalyticARB} and that the $NA(\mathcal{Q})$ condition hold true. Let $H: \O^T \to \mathbb{R}$ be usa.  Then, there exists some $\widehat P \in \mathfrak{P}(\O^{T})$ such that NA$(\widehat P)$ holds true, $D_{\widehat{P}}^{t+1}(\cdot)=D_{\mathcal{Q}}^{t+1}(\cdot)$ for all $0\leq t \leq T-1$ 
and 
\begin{small}
\begin{align}
\label{pricetilde}
-\infty<\pi^{\mathcal{Q}}(H)=\pi^{\widetilde{\mathcal{P}}}(H)=\underline{\pi}^{\widetilde{\mathcal{P}}}(H)=\sup_{P \in \widetilde{\mathcal{P}}} \pi^{P}(H)=\pi^{\widehat P}(H)=\sup_{R \in \mathcal{M}_{e,\widetilde{\mathcal{P}}}} E_{R} (H)=\sup_{R \in \mathcal{M}_{a,\Qc}} E_{R} (H), 
\end{align}
\end{small}
where $\widetilde{\mathcal{P}}$ is defined in \eqref{PstarARB}. 
\end{theorem}
\begin{proof}
This is an application of Theorem \ref{super} together with  Proposition \ref{Att}. In order to get a full support probability measure, we define $\widehat{P}$ slightly differently than in Theorem \ref{super}.  
Let $(P_n)_{n \geq 1} \subset \widetilde{\mathcal{P}}$  such that $\sup_{P \in \widetilde{\mathcal{P}}}\pi^{P}(H)=\sup_{n \geq 1}\pi^{P_n}(H)$. 
For all $n \geq 1,$  $P_n=P^n_1\otimes  p^n_2 \otimes \ldots \otimes p^n_T $ where $P_1^n \in \widetilde{\mathcal{P}}^{1}$ and $p^n_{t+1}(\cdot,\o^{t}) \in \widetilde{\mathcal{P}}_{t+1}(\o^{t})$ for all $\o^{t} \in \O^{t}$  and for all $1\leq t \leq T-1.$ 
Let $\widehat{P}_1 :=\sum_{n \geq 1} 2^{-n}P^n_1$ and $\widehat{p}_{t+1} :=\sum_{n \geq 1} 2^{-n}{p}^n_{t+1},$ for all $1\leq t \leq T-1.$ 
Then, we set $\widehat P:=P_1\otimes  p_2 \otimes \ldots \otimes p_T.$ Using Fubini theorem, for all $A \in \Bc(\O^T)$
$$ \widehat P(A)= \sum_{n_T \geq 1} 2^{-n_T} \ldots \sum_{n_1 \geq1} 2^{-n_1} \int_{\O^T}1_A(\o^T) p_T^{n_T}(d\o^T,\o_{T-1}) \ldots P^{n_1}_1(d\o^1).$$
So, for all $n \geq 1$, $P_n \ll  \widehat{P}.$ If now $A$ is a $\widetilde{\mathcal{P}}$ polar set, $P^{n_1}_1\otimes  p^{n_2}_2 \otimes \ldots \otimes p^{n_T}_T(A)=0$ and thus, $A$ is a $\widehat{P}$-null set. So, as in the proof of Theorem \ref{super}, it follows that 
 $\pi^{\widetilde{\mathcal{P}}}(H)  =\pi^{\widehat{P}}(H)$ and NA($\widehat{P}$) holds true. \\
As $P_n \in \widetilde{\mathcal{P}},$  for all $0\leq t \leq T-1,$
\eqref{moon} shows that $D_{P_n}^{t+1}(\cdot)=D_{\mathcal{Q}}^{t+1}(\cdot),$  
$\widehat{p}_{t+1} (D_{\mathcal{Q}}^{t+1}(\o^t),\o^t)=1,$ for all $\o^t \in \O^t$ and  thus  Lemma \ref{supportcar} shows that  $D_{\widehat{P}}^{t+1}(\cdot) \subset D_{\mathcal{Q}}^{t+1}(\cdot).$\\
Fix  $0\leq t \leq T-1$ and $\o^t \in \O^t$. Then we have that $\sum_{n \geq 1} 2^{-n}{p}^n_{t+1}(D_{\widehat{P}}^{t+1}(\o^t),\o^t)=1$ and so  ${p}^n_{t+1}(D_{\widehat{P}}^{t+1}(\o^t),\o^t)=1$ for all $n\geq 1$. Thus, $D_{\mathcal{Q}}^{t+1}(\cdot)=D_{P_n}^{t+1}(\cdot) \subset D_{\widehat{P}}^{t+1}(\cdot).$  
\end{proof}

\begin{example}
\label{exN}
We illustrate  why without further assumption, one cannot get $\widehat{P} \in \mathcal{P}$ 
 in Theorem \ref{super} even in the dominated case. Suppose that $T = 1$, $d=1$ and   $\Omega=\mathbb{R}$ % Assume that the risk free asset has a price  constant equal to $1$ and that 
The (discounted) risky asset is such that   $S_0=0$ and $S_{1}(\o)=\o,$ for all $\o \in \mathbb{R}.$ 
Let $H=1_{S \geq 1}$.  
For $n \geq 1$, we set $P_{n}=\frac{1}{2}\delta_{\{-1\}}+\frac{1}{2}\delta_{\{1+\frac{1}{n}\}}$, $\mathcal{Q}=\mathcal{P}:=\mbox{conv} \{P_{n}, n \geq 1\}$. It is clear that $NA(P_{n})$ holds true for all $n\geq1$, so $sNA(\mathcal{Q})$ and $NA(\mathcal{Q})$ also hold true.
For all $n\geq 1$,  $D_{P_{n}}=\{{-1}\} \cup \{1+\frac{1}{n}\}$ and  $D_{\mathcal{Q}}=\{-1\}\cup \{1\} \cup \{1+\frac{1}{n}, n\geq 1\}$. Finally, note that $\pi^{P_{n}}(H)=\frac{n}{2n+1}$ and $\pi^{\mathcal{Q}}(H)=\frac{1}{2}$.
It is clear that for all $P \in \mathcal{Q}$, $\pi^{P}(H)<\frac{1}{2},$ so one cannot hope to find $\widehat{P} \in \mathcal{Q}$ such that $\pi^{\widehat{P}}(H)=\frac{1}{2}$.  Similarly, $1 \notin D_{P}$, so  one cannot hope to find $\widetilde{P} \in \mathcal{Q}$ such that $D_{\mathcal{Q}}=D_{\widetilde{P}}$ in Lemma \ref{lemmptilde}. 
%For all $P \in \mathcal{Q}$, there exists some $m\geq$ such that $P(S_{1}(\cdot) \in ]\frac{1}{2},1+\frac{1}{m}[=0$, so $D_{P} \cap [\frac{1}{2},1+\frac{1}{m}]=\emptyset$ and $p^{P}$
\end{example}
\begin{remark}
Note that recalling \eqref{Phat} and \eqref{PstarM},   if we assume that for all $1\leq t\leq T-1$,  the sets $\mathcal{Q}_{t+1}$  are countably convex (i.e. stable by countable convex combinations),
then $\widehat{P} \in \mathcal{P}$ in Theorem \ref{super} and $\widetilde{P} \in \mathcal{Q}$ in Lemma \ref{lemmptilde}. Thus, in Theorem \ref{theo2}
$\widetilde{\mathcal{P}} \subset \mathcal{Q}$ and  also $\widehat{P} \in  \widehat{P}.$

\end{remark}

%\section*{Appendix}

%\bibliographystyle{siamplain}
\bibliography{biblioRomainL}

\end{document}